\documentclass[sigconf]{acmart}

\acmSubmissionID{1380}

\usepackage{float}
\usepackage{placeins}
\usepackage{xcolor}
\usepackage{comment}
\usepackage{url}
\usepackage{amsmath}
\usepackage{caption}
\usepackage{subcaption}

\usepackage{microtype}

\usepackage{array}

 \usepackage{verbatim}

\usepackage[ruled,linesnumbered]{algorithm2e}
\SetKwComment{Comment}{/* }{ */}

\title{Recursive Methods for Synthesizing Permutations on Limited-Connectivity Quantum Computers}

\begin{document}


\author{Cynthia Chen}
\affiliation{%
  \institution{Caltech}
  \city{Pasadena}
  \country{USA}}
\email{cchen6@caltech.edu}

\author{Bruno Schmitt}
\affiliation{%
  \institution{EPFL}
  \city{Lausanne}
  \country{Switzerland}}
\email{bruno.schmitt@epfl.ch}

\author{Helena Zhang}
\affiliation{%
  \institution{IBM Quantum
  }
  \city{Yorktown Heights, NY}
  \country{USA}}
\email{helena.zhang@ibm.com}

\author{Lev S. Bishop}
\affiliation{%
  \institution{IBM Quantum
  }
  \city{Yorktown Heights, NY}
  \country{USA}}
\email{lsbishop@us.ibm.com}

\author{Ali Javadi-Abhari}
\affiliation{%
  \institution{IBM Quantum
  }
  \city{Yorktown Heights, NY}
  \country{USA}}
\email{ali.javadi@ibm.com}

\begin{abstract}
  We describe a family of recursive methods for the synthesis of qubit permutations on quantum computers with limited qubit connectivity. Two objectives are of importance: circuit size and depth. In each case we combine a scalable heuristic with a non-scalable, yet exact, synthesis. Our algorithms are applicable to generic connectivity constraints, scale favorably, and achieve close-to-optimal performance in many cases. We demonstrate the utility of these algorithms by optimizing the compilation of Quantum Volume circuits, and to disprove an old conjecture on reversals being the hardest permutation on a path.
\end{abstract}

\settopmatter{printacmref=false}
\setcopyright{none}
\renewcommand\footnotetextcopyrightpermission[1]{}
\pagestyle{plain}

\maketitle

\section{Introduction}\label{sec:intro}
There is a strong belief that quantum computers will be able to solve certain problems beyond the reach of classical computers~\cite{vazirani, shor}. Typically, researchers describe quantum algorithms in greatly abstracted terms, whether pure mathematical representations or high-level quantum circuits---a sequence of operators acting on qubits.  While these high-level circuits assume all-to-all connectivity, in practice, the qubits in most quantum hardware are not fully connected due to noise and interconnect challenges. Therefore, not every qubit pair can participate in the same gate operation. These connectivity restrictions are known as coupling constraints.

The task of finding a mapping from virtual instructions to allowed physical instructions is known as quantum circuit mapping. Completing this task is not always possible without applying additional operators to the circuit. These additional operators enable the execution of gates on non-adjacent qubits by permuting their place in the coupling graph. Their cost, however, can dominate the total cost for many applications. This work focuses on families of circuits that \emph{permute} qubits on a coupling graph. Also, these permutation circuits appear prominently in quantum computing benchmarks~\cite{Cross}.

The size and depth of a circuit are two important metrics for evaluating the quality of a synthesis process. Size refers to the number of gates. Since each gate is noisy, many gates can cause errors to accumulate. Depth refers to the number of timesteps (or layers) in the circuit. Since qubits have limited coherence, they retain information for a short time. Hence, deep circuits can get overwhelmed by noise. Note that there is often a tradeoff between size and depth optimization.

An intuitive procedure for the physical synthesis of permutations is to perform SWAP operations between pairs of qubits. The size-optimizing version of this problem is known as token swapping~\cite{tokenswapping}, and the depth-optimizing version is known as routing via matchings~\cite{Alon}. Finding optimal solutions to either of these problems takes exponential time. While swapping qubits is the most common model employed to solve this problem due to its analogy to graph algorithms, we can improve upon this by leveraging fundamental quantum gates such as CNOT. Since permutations are a special class of linear functions over $\mathbb{F}_2$, they are computable using a series of XOR operations (i.e., CNOTs).

Broadly, existing methods for synthesizing permutation circuits either optimize for size or depth, using either SWAPs or CNOTs. Some methods are exact but non-scalable, some are tailored to certain topologies, and some are general-purpose heuristics. In Tables~\ref{tab:size} and~\ref{tab:depth} we summarize prior literature, ordered chronologically and separated by whether they optimize for size or depth.



\begin{table*}
\begin{center}
\begin{tabular}{  m{3cm }| m{5cm}| m{2cm}| m{3.5cm} |  m{2.5cm}|}

 & Description & Primitive Op. & Topologies & Size bounds \\
\hline
Patel ('08)~\cite{patel} & Grouped Gaussian elimination & CNOT & Fully connected & $O\left(\frac{n^2}{\log n}\right)$\\
\hline
Kissinger ('19)~\cite{Kissinger},\hspace{.5cm} Gheorghiu ('20)~\cite{Gheorghiu} & Gaussian elimination with Steiner trees & CNOT & Any & $O(n^2)$\\
\hline
Wu ('19)~\cite{Wu} & Recursive elimination of $ROW_i$/$COL_i$ & CNOT & Any & $2n^2$\\
\hline
Childs ('19)~\cite{childs2019circuit} & token swapping approximation & SWAP & Any &  $4 \times$ optimal \\
\hline
Schmitt ('20)~\cite{Schmitt} & A* search with admissible heuristic & SWAP & Any & Optimal (small $n$)\\
\hline
{\bf Ours} & SAT formulation & CNOT & Any & Optimal (small $n$)\\
\hline
\end{tabular}

\end{center}
\caption{Summary of size-optimizing methods. Bounds are in terms of the operation the method is based on. $1$ SWAP = $3$ CNOTs.}
\label{tab:size}
\end{table*}

\begin{table*}
\begin{center}
\begin{tabular}{  m{3cm }| m{5cm}| m{2cm}| m{3.5cm} |  m{2.5cm}|}
 & Description & Primitive Op. & Topologies & Depth bounds \\
\hline
Alon ('94)\cite{Alon} & Graph matchings & SWAP & Tree, Cartesian prod., etc.
& $3n$ \\
\hline
Zhang ('97) \cite{Zhang} & Caterpillar partition and matchings. 
& SWAP & Tree & $\frac{3}{2}n + O(\log n)$\\
\hline
Kutin ('07) \cite{Kutin} & Odd-even transposition sort (+ manual for specific permutations) & CNOT & Line & $n$\\
\hline
Maslov ('08) \cite{maslov} & Divide and conquer & SWAP & Any & $8n + const$ \\
\hline
Wu ('19) \cite{Wu} & Using $n^2$ ancillas & CNOT & 2D grid & $O(\log n)$\\
\hline
Schmitt ('20)~\cite{Schmitt} & SAT formulation & SWAP & Any & Optimal (small $n$)\\
\hline
de Brugiere ('21) \cite{Brugiere} & Divide and conquer & CNOT& Fully connected & $\frac{4}{3}n + 8 \log_2(n)$\\
\hline
Bapat ('21) \cite{Bapat} & Divide and conquer & Reversal & Any & $O(k^2) + \frac{2}{3}r$
\\
\hline
{\bf Ours} & SAT formulation & CNOT & Any & Optimal (small $n$)\\
\hline
{\bf Ours} & Divide and conquer & SWAP & Any & $2n + 2\log{n}$\\
\hline
\end{tabular}
\end{center}
\caption{Summary of depth-optimizing methods. Bounds are in terms of operation the method is based on. $1$ SWAP = $3$ CNOTs.}
\label{tab:depth}
\end{table*}

\subsection{Our Contributions}

\begin{figure}[t]
\centering
\includegraphics[width=0.92\columnwidth]{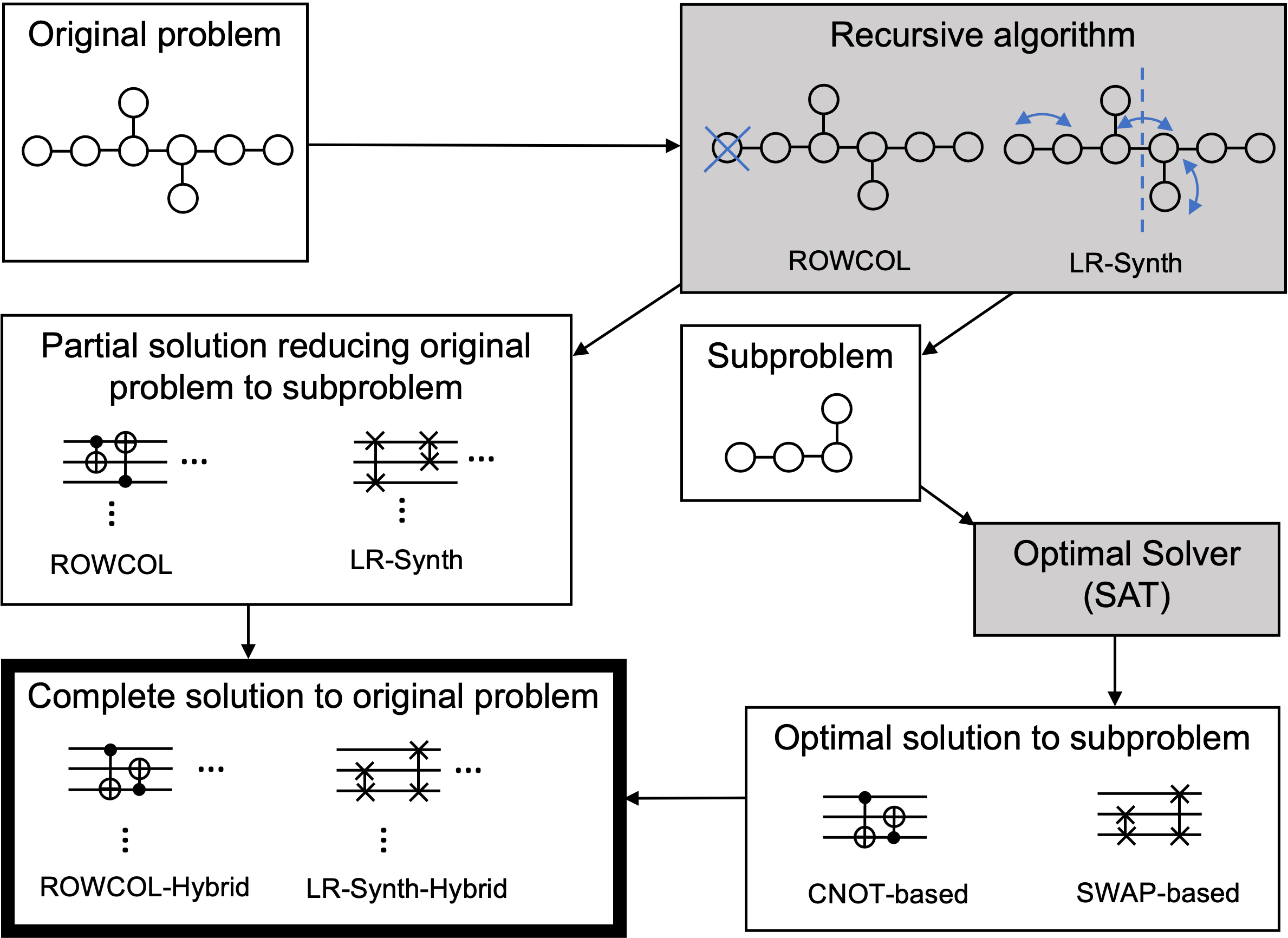}
\caption{High-level view of our hybrid approach for synthesizing permutations on arbitrary topologies. It takes as input a graph and the desired permutation, and uses ROWCOL or LR-Synth heuristics to optimize for size or depth, respectively. Both generate a partial solution that is combined with an optimal solution synthesized by a SAT-based technique.}
\label{fig:overview}
\end{figure}

In this work, we utilize recursive heuristics for the synthesis of permutations. These have two key advantages: first, they reduce the problem at each stage, making high-quality solutions easier to find and can be used in conjunction with optimal methods in the inner stages of recursion (Figure~\ref{fig:overview}). Second, with depth optimization, it is important to parallelize the circuit as much as possible, and recursive methods can be parallelized at each stage. Our approach scales to thousands of qubits, applies to any qubit connectivity topology, and yields better results than state-of-the-art approaches.

First, we propose a SAT encoding of the problem, which can find circuits with optimal size or depth using CNOTs. This technique improves over prior SWAP-based optimal solvers~\cite{Schmitt} and can be used to synthesize any linear function, not just permutations.

Equipped with optimal solvers (hereafter referred to as {\em CNOT-size-optimal}, {\em CNOT-depth-optimal}, {\em SWAP-size-optimal}, {\em SWAP-depth-optimal}), we use them in recursive heuristics and scale to much larger problem instances. For size, we rely on a modified version of the ROWCOL heuristic by Wu et al.~\cite{Wu}, which outperforms other size-oriented algorithms and lends itself to recursion. No such algorithm exists for depth, so we propose {\em LR-Synth}, a novel depth-optimizing divide-and-conquer heuristic for general graphs.

ROWCOL is a recursive algorithm that reduces the problem by removing one non-cut vertex/qubit from the connectivity graph at a time. We observe that the order of vertex elimination is important and optimize for that. Combining ROWCOL with CNOT-size-optimal for the inner stages of recursion yields our hybrid algorithm, which significantly outperforms prior work.

Our second recursive algorithm, LR-Synth, is a novel algorithm that routes qubits to the correct half of the topology in each step then recurses on the left and right subgraphs of the topology. Since we can always divide a topology into two parts, an advantage of LR-Synth is that it can be applied to any topology, which is not the case for many tailored prior approaches. We benchmark LR-Synth against Schmitt's SAT-based token swapping solver~\cite{Schmitt}, which solves for optimal SWAP depth. We also benchmark LR-Synth against state-of-the-art heuristics that exist for specific topologies (paths, trees, and lattices). Overall we find that LR-Synth achieves close to optimal circuit depth while being much more scalable than exact solvers and applicable to any topology. 

We note that there are two gaps in prior work which we fill. First, while exact solvers exist for size-optimal and depth-optimal synthesis based on SWAPs~\cite{Schmitt} (and available in software package Tweedledum~\cite{tweedledum}), no CNOT-optimal synthesizers are known. Since each SWAP is equivalent to three CNOTs,  direct usage of CNOTs in synthesis can yield better circuits. 

Second, we propose a depth minimization technique on general topologies using SWAPs. 
Compared to other depth optimizing heuristics on general topologies, our method takes advantage of more parallel routing opportunities than \cite{maslov} and is more general than \cite{Bapat}, which assumes the existence of fast reversal operations that may not be available on some architectures.

Finally, we use our exact algorithm to disprove a 15-year-old conjecture by Kutin et al.~\cite{Kutin} that reversal is at least as depth-intensive to synthesize with CNOTs as any other permutation on a path.

\section{Optimal SAT-based algorithm for CNOT circuits}\label{sec:optimal}

We formulate the problem of finding an optimal-depth circuit for a linear matrix representing a reversible function as instances of the Boolean satisfiability problem. Permutations are special cases of linear functions, and this method yields direct synthesis of permutations using CNOTs. In our encoding, we use two kinds of variables. Matrix variables, $m^{d}_{i, k}$, which indicate whether a matrix entry $(i, k)$ is $0$ or $1$ at depth $d$; and CNOT gate variables, $g^{d}_{c \to t}$, which indicate that a CNOT between qubits $c$ and $t$ took place at depth $d$. For example a 3-qubit linear function synthesis would be encoded as such, where a $3x3$ Boolean matrix represents the linear function over 3 bits, and a CNOT application transforms the matrix by XOR-ing two rows.

\[
    \small
    \begin{bmatrix}
        m^{d}_{0, 0} & m^{d}_{0, 1} & m^{d}_{0, 2}\\
        m^{d}_{1, 0} & m^{d}_{1, 1} & m^{d}_{1, 2}\\
        m^{d}_{2, 0} & m^{d}_{2, 1} & m^{d}_{2, 2}\\
    \end{bmatrix}
    \xrightarrow[g^{d}_{0 \to 2}]{}
    \begin{bmatrix}
        m^{d}_{0, 0} & m^{d}_{0, 1} & m^{d}_{0, 2}\\
        m^{d}_{1, 0} & m^{d}_{1, 1} & m^{d}_{1, 2}\\
        m^{d}_{0, 0} \oplus m^{d}_{2, 0}  & m^{d}_{0, 1} \oplus m^{d}_{2, 1} & m^{d}_{0, 2} \oplus m^{d}_{2, 2}\\
    \end{bmatrix}
\]

Our encoding uses four different types of clauses to constrain the problem such that the solution corresponds to a valid linear reversible circuit:

\begin{itemize}
    \item{\textbf{C1.}} Each depth that does not hold the target transformation must have at least one CNOT.
    \[ \forall\{d : \exists(d+1)\},\; \sum_{(c, t) \in E} g^d_{c\to t} + g^d_{t\to c} \geq 1. \]
    
    \item{\textbf{C2.}} At each depth that has at least one CNOT, each qubit can only be involved in one CNOT.
    \[ \forall\{d : \exists(d+1)\},\, \forall{c} \in V,\; \sum_{t \in \delta(c)} g^d_{c\to t} + g^d_{t\to c} = 1.\]
    where $\delta(c)$ is the set of qubits adjacent to $c$.
    
    \item{\textbf{C3.}} If at depth $d$ the variable indicating a $CNOT(i, k)$ is true, then all elements of the $k$-th row at depth $d+1$ must be XOR-ed between the element and its corresponding element in the $i$-th row at the previous depth $d$.
    \begin{equation*}
        g^d_{c\to t} \implies \bigwedge_j \left( m^{d+1}_{t, j} = m^{d}_{t, j} \oplus m^{d}_{c, j} \right)
    \end{equation*}

    \item{\textbf{C4.}} If at depths $d$ and $d+1$ a matrix entry in the $i$-th row has different values, then exactly one of the CNOT variables that has $i$ as target must be true.
    \begin{equation*}
        m^{d+1}_{t, j} \neq m^{d}_{t, j} \implies \sum_{c \in \delta(t)} g^d_{c\to t} = 1
    \end{equation*}
\end{itemize}

The SAT solver only answers whether a given formula is satisfiable or unsatisfiable. Therefore, we need to translate our optimization objective into a series of queries to the SAT solver. In this case, each query ``asks'' the solver if there exists a circuit that implements the desired transformation using a specified depth. Our implementation incrementally solves the problem: first, we build a formula that encodes a solution with a specified depth using the above constraints; then, if the formula is unsatisfiable, we increment the depth by adding new variables and constraints. We keep incrementing the depth until we find a satisfiable formula to decode and build a linear reversible circuit.

We can use a slightly different encoding to find reversible CNOT circuits with optimal size. In such a case, the only difference lies in the first type of constraints: Instead of requiring depths to have at least one CNOT, we restrict the number to exactly one.

\begin{figure}[htp]
\centering
\includegraphics[width=0.9\columnwidth]{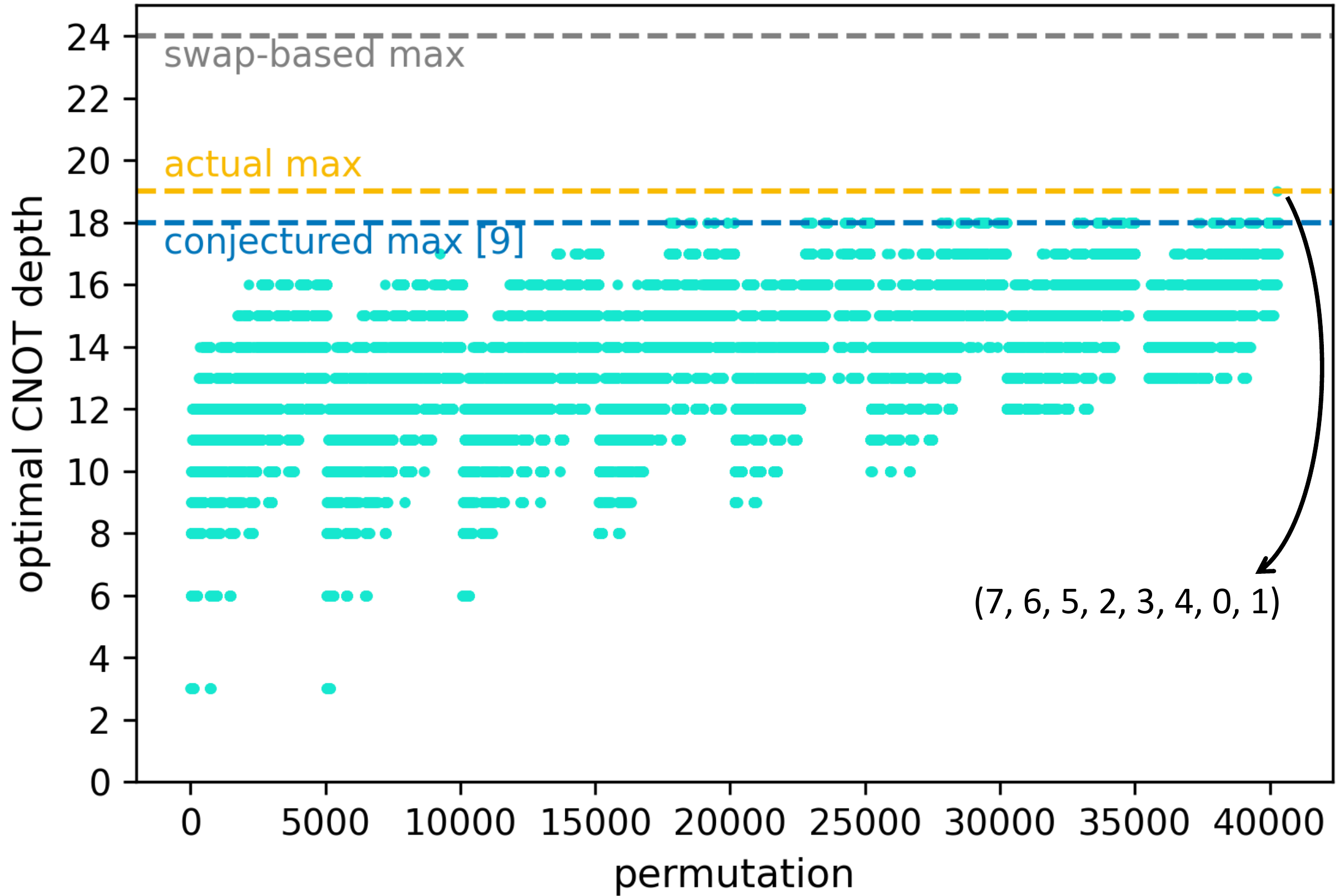}
\caption{Optimal depth of all permutations on an 8-qubit path synthesized by CNOT-depth-optimal. A specific permutation is shown that has depth higher than the max conjectured by~\cite{Kutin}. The best max under a SWAP-based synthesis is also shown for comparison. 
96\% of permutations have better depth using CNOTs as opposed to SWAPs.
}
\label{fig:cnot-depth-optimal}
\end{figure}

As we will show, these SAT-based solutions can be applicable with scalable heuristics to improve their quality, but, they are also useful on their own. For example, we studied CNOT-depth-optimal solutions for permutations on a path, which was the focus of Kutin et al.'s paper~\cite{Kutin}. They conjectured that reversals are at least as hard as any other permutation, and that reversals are synthesizable with depth $2n+2$ CNOTs. However, by solving for all instances of 8-qubit permutations, we found a specific permutation that required depth $2n+3=19$, as shown in Figure~\ref{fig:cnot-depth-optimal}, thus disproving the conjecture.

\section{Size Optimization: ROWCOL-Hybrid}\label{sec:size}

\subsection{Algorithm Description}
We propose and implement a hybrid CNOT circuit synthesis algorithm that combines a modification on Wu's ROWCOL algorithm~\cite{Wu} with optimal SAT-based methods (Figure \ref{fig:overview}).

\begin{figure}[htp]
\centering
\includegraphics[width=1\columnwidth]{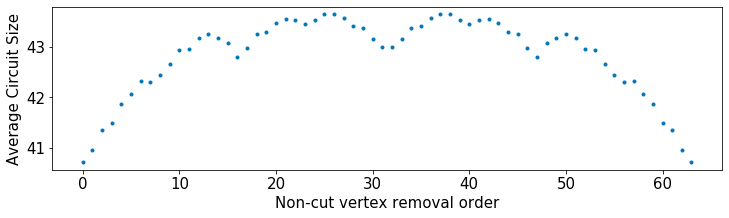}
\caption{Effect of vertex removal order on average circuit size synthesized by ROWCOL for an 8-qubit path.}
\label{fig:orders}
\end{figure}

In \cite{Wu}, non-cut vertices are iteratively removed from the graph, though choosing the order of removing non-cut vertices is not discussed. We find that removal order has a non-trivial effect on circuit size and depth (Figure \ref{fig:orders}). We run Wu's algorithm for all possible non-cut vertex orders to reduce circuit size. For each permutation, select an optimal ordering, defined as one that results in the smallest CNOT circuit size, with depth as a tie-breaker.
\footnote{Step 6 in Wu's ROWCOL algorithm has to be brute-forced, which increases algorithm runtime.}

Since the ROWCOL algorithm reduces the problem size by one qubit in each iteration, it lends itself to hybridization with exact solvers. We thus terminate the heuristic early when the graph has only four qubits left (threshold chosen empirically), then call our CNOT-size-optimal solver on the reduced graph to finish synthesizing the circuit. (Note that SWAP-size-optimal is not an option since the sub-problems may not be permutations). We combine the circuits obtained by both methods to obtain the final result.


\subsection{Results}
\begin{figure}[htp]
\centering
\includegraphics[width=0.92\columnwidth]{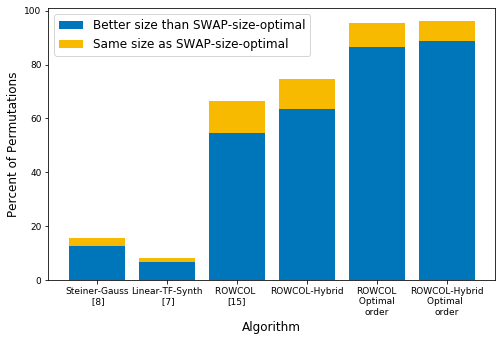}
\caption{Comparison of CNOT-size-optimizing methods and how much they improve upon SWAP-size-optimal methods (permutations on a path of 8). Our hybrid with optimal ordering approach achieves smaller CNOT count for 88.8\% of all permutations.
}
\label{fig:algComp}
\end{figure}

We compare our approach to three previous CNOT-based methods: ``Steiner-Gauss" by Kissinger et al.~\cite{Kissinger}, ``Linear-TF-Synth" by Gheorghiu et al.~\cite{Gheorghiu} and ``ROWCOL" by Wu et al.~\cite{Wu}, over all 8-qubit permutations on a path topology, as shown in Figure \ref{fig:algComp}. Our ROWCOL-Hybrid algorithm with optimal vertex order achieves a smaller CNOT count than the size-optimal SWAP-based method for 88.8\% of all permutations. At the same time, Kissinger et al.~\cite{Kissinger}, Wu et al.~\cite{Wu}, and Gheorghiu et al.~\cite{Gheorghiu} synthesize smaller CNOT circuits than the optimal SWAP-based method for 12.5\%, 54.8\%, and 6.6\% of all permutations, respectively.

\begin{figure*}[htp]
    \centering
    \includegraphics[width = 2\columnwidth]{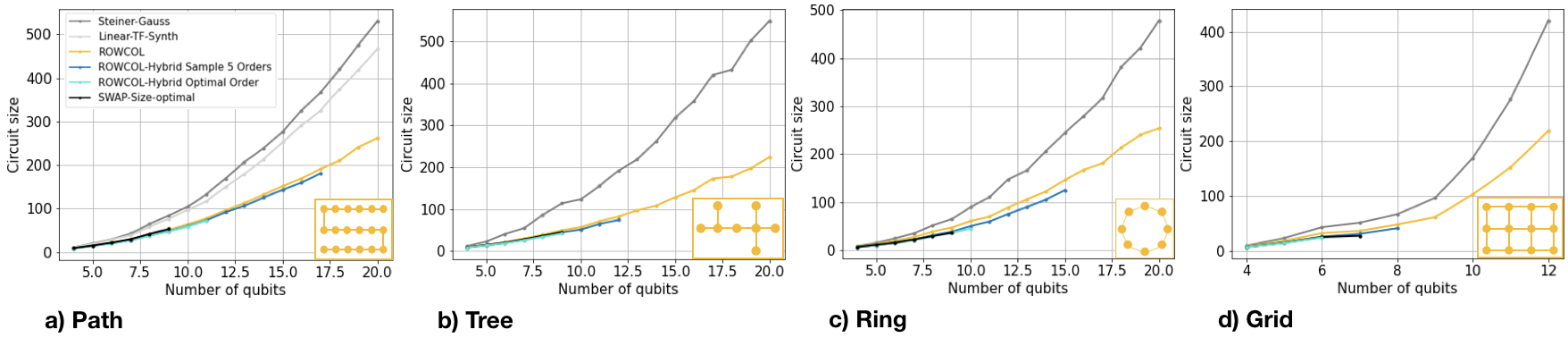}

    \caption{Size comparison of four topologies for permutations synthesized using ROWCOL-Hybrid (optimal vertex removal order and picking the best of five orders), SWAP-size-optimal, and existing heuristics.}
    \label{fig:sizeResults}
\end{figure*}

We also compare our approach to existing CNOT-based methods for randomly sampled permutations on increasing qubit numbers, as shown in Figure \ref{fig:sizeResults}. While ROWCOL-Hybrid is more computationally expensive than the existing heuristics, it scales better than the SAT solver and achieves the smallest average circuit sizes. 

We find this CNOT-based approach to be a surprisingly large improvement over using SWAPs, which naively seem to be naturally suited for permutations. While our approach primarily optimizes for circuit size, there is also an improvement in circuit depth compared to the three other algorithms.


\section{Depth Optimization: LR-Synth}\label{sec:depth}

Due to low qubit coherence times, circuit depth is often the most limiting factor in near-term quantum experiments. Existing depth-optimizing synthesis methods scale exponentially~\cite{Schmitt}, assume full connectivity~\cite{Brugiere}, are restricted to certain limited-connectivity topologies~\cite{Alon,Kutin,Zhang}, or assume non-standard primitive operations~\cite{Bapat}. We propose a polynomial-time divide-and-conquer algorithm, LR-Synth, for synthesizing permutations on any limited-connectivity topology, optimizing for depth of SWAPs. Our algorithm may also be of independent interest for general network routing.

\subsection{Algorithm Description}
\begin{figure}[H]
\centering
\includegraphics[width=0.90\columnwidth]{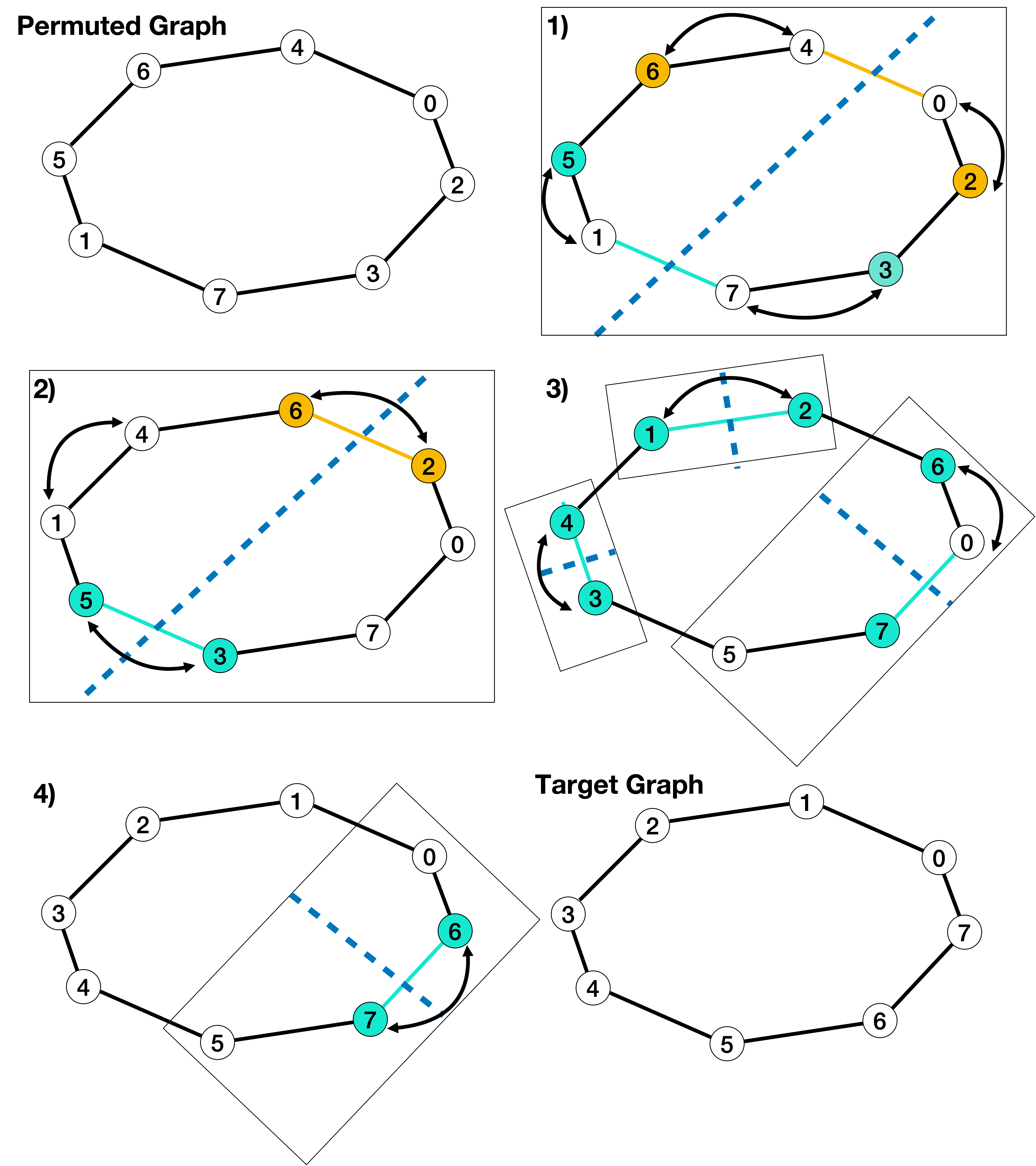}
\caption{Algorithm steps on an 8-qubit ring topology. In steps 1 and 2, there are two possible paths to move a qubit from one side to another (yellow and blue edges). Qubits that must be routed to the other side have the same color as the path they are assigned, and the algorithm recurses in steps 3 and 4 until the target graph is achieved.} 
\label{fig:divideConquer}
\end{figure}

Given a permuted graph $G_p$ and a target graph $G_t$, the goal is to convert $G_p$ to $G_t$ through a series of SWAPs, parallelizing as many SWAPs as possible. The high-level idea is as follows (see Figure~\ref{fig:divideConquer}): We partition $G_p$ into two connected subgraphs as close in size as possible, then we move qubits to the correct half using maximal matchings. Once all qubits are on the correct half, we recursively call the algorithm on each half. Since the left and right halves are disjoint, their circuitry is parallelized on chip. The complete algorithm is given in Algorithm~\ref{alg:LR-Synth}, where $D(G, a, b)$ denotes the shortest distance between vertices $a$ and $b$ in $G$. 

The first step is partitioning $G_p$ into two connected subgraphs, $leftG$ and $rightG$, as balanced as possible (step \ref{op:a}). This is the balanced connected 2-partition problem and is NP-hard~\cite{CHLEBIKOVA}, so we use a heuristic to perform the partition. 
We use a simple heuristic that performs a depth first search starting from each node until half of the graph is traversed, removing the visited node each time to put into $leftG$, while keeping $rightG$---the part of the graph not traversed---connected. We find that this heuristic finds adequate splits for most topologies of interest. To maximize parallel routing across the partition, we select partitions resulting from removing vertex disjoint edges from $G$ if possible (let $removedEdges$ be all edges in $G_p$ but not in $leftG$ or $rightG$), because if two or more edges share a vertex, that vertex would become the bottleneck for routing vertices across the partition. In cases where no such partitions exist, we let $removedEdges$ be edges that form a maximal matching. Since trying every partition is infeasible for larger topologies, we sample up to $S$ partitions.

Next, we assign a path to each vertex that needs to be moved to the other half, specified by one of the removed edges (step \ref{op:b}). There are many possible heuristics for $assignPath$. Our heuristic assigns $v \in moveToRight$ to a path by choosing the $(l, r)$ that minimizes 
\begin{equation*}
\begin{split}
Cost = &\max \{D(leftG, v, l), D(rightG, w, r)\}  + n_{(l, r)}/2,
\end{split}
\end{equation*}
where $w$ is the unassigned vertex in $moveToLeft$ closest to $(l, r)$ and $n_{(l, r)}$ is the number of nodes assigned to $(l, r)$. $D(leftG, v, l)$ and $D(rightG, w, r)$ are the number of swaps needed to move $v$ and $w$ respectively to the removed edge. Since this can be performed in parallel, we take the maximum distance. We add the term $n_{(l, r)}$ because $v$ and $w$ cannot be routed to the other side until qubits already assigned to the path $(l, r)$ are routed, so $n_{(l, r)}$ penalizes moving too many qubits via one path. 

After assigning paths, we iterate and add swaps until all qubits are routed to the correct side of the topology via their assigned paths (step \ref{op:z}). Let $E$ be a set of potential edges to swap in a given iteration. At each iteration, we prioritize adding a swap by assigning it a weight of $1.3$ if the swap moves vertex $v$ to its final destination and all vertices on the path from $v$ to a terminal node of $G$ are already correctly positioned because making the swap effectively reduces the size of the topology we work with (step \ref{op:c}).  

To explain steps \ref{op:h} - \ref{op:i}, we consider, without loss of generality, $u\in moveToRight$, where $u$ is routed across the removed edge $(l, r)$. For any vertex $u \in moveToRight$, we would want to add a potential SWAP between $u$ and $v$ when  $D(G, u, r) > D(G, v, r)$ (step \ref{op:d}), since this would bring $u$ closer to the $rightG$. If $v\in moveToRight$ and $vertexToPath[v] = (l, r)$ and $D(G_t, u, rT) \leq D(G_t, rT)$, $u$ is closer to the removed edge in the target graph, implying it should be routed to the right after $v$ is routed, so we do not add $(u, v)$ to $E$ in this iteration (step \ref{op:e}).  If no possible SWAPs are present in a given iteration, step \ref{op:jj} fixes this in the next iteration by swapping $u$ and $v$ if there is a neighbor $b$ of $v$ that is not in $moveToRight$ and $b$ is closer than $u$ to $r$, since then in a subsequent iteration, swapping $u$ and $b$ would move $u$ closer to its target location. 

If $v \in moveToRight$ and $v$'s path is different from $u$, then we add $(u, v)$ to $E$ if the swap brings both $u$ and $v$ closer to their respective paths (step \ref{op:y}). If $v$ is already on the correct side in $leftG$, then we add $(u, v)$ to $E$ because this would bring $u$ closer to its destination without moving $v$ to the wrong side (step \ref{op:f}). Finally, we swap $u$ to the right side if $v \in moveToLeft$; otherwise, extra swaps would be needed to move $v$ back to $rightG$ in a future iteration. We prioritize swaps that bring vertices to the correct side since this would allow more vertices to be routed via the same path, so we weight these swaps by $1.2$ (step \ref{op:g}). 

After steps \ref{op:h} - \ref{op:i},  we find a maxMatching to minimize depth (step \ref{op:j}). In step \ref{op:k}, if $G$ is a path or a ring topology, for each edge $(u, v)$ in $G$, if $u$ and $v$ are not in $moveToLeft$ or  $moveToRight$ and their orders are flipped in $G_t$, we add $(u, v)$ to $maxMatching$ if adding the edge still results in a matching. This adds swaps that would otherwise occur later in the algorithm and takes advantage of earlier matchings to reduce depth.

Once $moveToRight$ and $moveToLeft$ are emptied, all vertices are on the correct half of the topology. After sampling $S$ partitions, we choose the partition that empties  $moveToRight$ and $moveToLeft$ in the least number of iterations (lowest depth) and use how close $G$'s state is to $G_t$ as a tiebreaker (step \ref{op:l}). Finally, we recursively call the algorithm on $leftG$ and $rightG$ (steps \ref{op:o}, \ref{op:p}). Since $leftG$ and $rightG$ contain disjoint vertices, the two calls are parallelized.  

\SetAlFnt{\small}
\begin{algorithm}[htp]
\caption{LR-Synth}
\label{alg:LR-Synth}
\KwData{(i) Permuted graph $G_p$, (ii) Target graph $G_t$, (iii) Number of splits to sample S}
\KwResult{List of SWAPs taking $G_p$ to $G_t$}
leftGs, rightGs, removedEdges $\leftarrow$ partitionGraph\; \label{op:a}
Swaps $\leftarrow$ [\:]\; 
\For{split i = 1 to S} {
G $\leftarrow$ $G_p$.copy; leftG $\leftarrow$ leftGs[i]; rightG $\leftarrow$ rightGs[i]; removedEdges $\leftarrow$  removedEdges[i]\;
left$G_t$, right$G_t$ $\leftarrow$ corresponding left and right graphs of $G_t$\;
moveToLeft $\leftarrow$ \{v : v $\in$ rightG and v $\in$ left$G_t$\}; \;
moveToRight $\leftarrow$ \{v : v $\in$ leftG and  v $\in$ right$G_t$\}\;
vertexToPaths $\leftarrow$ assignPath\; \label{op:b}
GL $\leftarrow$ leftG $\cup$ removedEdges; GR $\leftarrow$ rightG $\cup$ removedEdges\; lockL $\leftarrow$ False; lockR $\leftarrow$ False\;

\While{moveToRight $+$ moveToLeft $\neq \emptyset$}{\label{op:z}
E = \{(u, v, 1.3) : (u, v) $\in$ G s.t. swapping u and v makes all vertices from $u$ to terminal of G in correct positions\}\;\label{op:c}
  \For{(u, v) $\in$ G s.t. u $\in$ moveToRight + moveToLeft}{\label{op:h}
  (l, r) $\leftarrow$ vertexToPaths[u]\; 
  (lT, rT) $\leftarrow$ corresponding edge in $G_t$\;
  
  \uIf{u $\in$ moveToRight and D(GL, u, r) $>$ D(GL, v, r)} { \label{op:d}
  
  \lIf {v $\in$ moveToRight and vertexToPath[v] = (l, r) and D($G_t$, u, rT) $\leq$ D($G_t$, v, rT)} {continue} \label{op:e}
  \lIf{lockL and $v \in$ moveToRight and vertexToPath[v] $\neq$ (l, r) and $\exists$ neighbor $b$ of $v$ s.t. $b \notin moveToRight$ and D(GL, u, r) $>$ D(GL, b, r)} 
  {E.add(u, v, 1); lockL $\leftarrow$ False} \label{op:jj}
  \uIf {v $\in$ moveToRight and vertexToPath[v] $\neq$ (l, r)}{\label{op:y}
  (a, b) $\leftarrow$ vertexToPath[v]\;
  \lIf {D(GL, v, b) $>$ D(GL, u, b)} {E.add(u, v, 1)}
  }
  \lElseIf{u $\neq$ l}{E.add(u, v, 1)} \label{op:f}
  \lElseIf{(u, v) = (l, r) and v in moveToLeft and vertexToPath[v] = (l, r)}{\label{op:g}
  E.add(u, v, 1.2)} \label{op:n}
  }
  
  \lIf{u $\in$ moveToLeft and D(GR, u, l) $>$ D(GR, v, l)} {
  Do similar logic as steps \ref{op:d} - \ref{op:n}
  }
  }\label{op:i}
  maxMatching $\leftarrow$ maximum weight matching for E\; \label{op:j}
  \lIf{maxMatching = $\emptyset$}{lockL $\leftarrow$ True; lockR $\leftarrow$ True}
  \lIf{G is a path or ring}{ \label{op:k}
  addToMatching}
  \lFor {(u, v) $\in$ maxMatching}{SWAP u and v in $G$}
  Update moveToLeft and moveToRight\;
}
Swaps $\leftarrow$ SWAPs from best split\label{op:l}\;
\lIf {size(leftG) > 1}{Swaps += LR-Synth(leftG, left$G_t$)} 
\label{op:o}
\lIf {size(rightG) > 1}{Swaps += LR-Synth(rightG, right$G_t$)} \label{op:p}
}
\Return Swaps
\end{algorithm}

\begin{figure*}[t]
    \centering
    \includegraphics[width = 2\columnwidth]{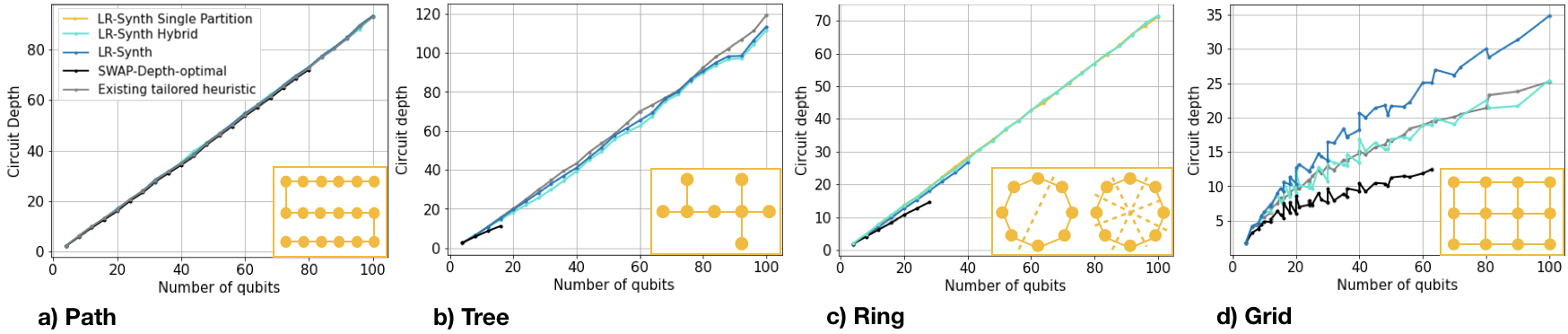}

    \caption{Depth comparison of four topologies for permutations synthesized using LR-Synth (hybrid, single, and all partitions), SWAP-depth-optimal, and tailored algorithms where one exists.}
    \label{fig:lr-synth}
\end{figure*}
\subsection{Results}
We benchmark our algorithm by comparing LR-Synth to Tweedledum's SWAP-depth-optimal SAT solver~\cite{tweedledum,Schmitt} for randomly sampled permutations on increasing qubit numbers. Due to the NP-completeness of the SAT problem, the solver does not terminate quickly for certain permutations even at relatively small qubit numbers, although the specific number is topology and permutation dependent.
To save computational time and resources, we run the SAT solver until the average circuit synthesis time exceeds 10 seconds on a typical personal computer. While LR-Synth can be applied to any limited connectivity topology, we perform our benchmarking on paths, trees, rings, and grids, commonly found on current physical devices. For line, tree and grid topologies, which have existing depth-optimizing heuristics, we also compare LR-Synth to them. For rings, there are many possible partitions for LR-Synth, so we compare the performance of randomly sampling a single partition versus selecting the best partition. 
We also hybridize LR-Synth with the SWAP-depth-optimal solver, analogous to the process for ROWCOL-Hybrid.

\subsubsection{Path}
We compare LR-Synth to Kutin's odd-even transposition sort algorithm and SWAP-depth-optimal by sampling 100 random permutations from 4-qubit to 100-qubit (Figure \ref{fig:lr-synth}a). We use Schoute's implementation of Kutin's algorithm~\cite{childs2019circuit, Schoute}. Kutin and LR-Synth achieve same sized circuits and similar depth circuits, which are better in average size than SWAP-Depth-optimal and slightly worse in average depth than SWAP-Depth-optimal. SWAP-Depth-optimal's average circuit synthesis time exceeded 10 seconds at 84Q.


\subsubsection{Tree}
We compare LR-synth to Zhang's tree algorithm and SWAP-depth-optimal by sampling 10 random trees for each tree size ranging from 4 to 100 qubits. 
For each tree topology, we sample 10 random permutations. We use Schoute's implementation of Zhang's algorithm \cite{Schoute}. SWAP-depth-optimal's average circuit synthesis time exceeds 10 seconds at 20 qubits. LR-Synth performs slightly better on average in terms of depth than Zhang.


\subsubsection{Ring}
We compare LR-Synth, sampling a single and all partitions, to SWAP-depth-optimal, averaging the results of 100 random permutations from 4-qubit to 100-qubit. The average circuit synthesis time of LR-Synth sampling all partitions exceeds 10 seconds at 44 qubits, compared to 28 qubits for SWAP-depth-optimal. The average circuit depths and sizes of sampling a single partition are comparable to those of all partitions, while the run time is significantly better.


\subsubsection{Grid}
Since grids have a well-defined structure, we partition $G_p$ into two subgraphs by performing a cut midway along the $x$ or $y$ direction. We benchmark LR-Synth against SWAP-Depth-optimal and Alon's Cartesian algorithm for grid sizes (2, 2) to (10, 10). For each grid size, we sample 100 random permutations.  Compared to other topologies, permutations on grids yield smaller average size and depth circuits. We observe this behavior for both LR-Synth and SWAP-depth-optimal, and it is an intuitive result because there are more paths to route qubits that can be parallelized. We also note that the hybrid algorithm sees significant performance improvements on this topology.

 \subsubsection{Runtime}
\begin{figure}[htp]
 \centering
 \includegraphics[width=\columnwidth]{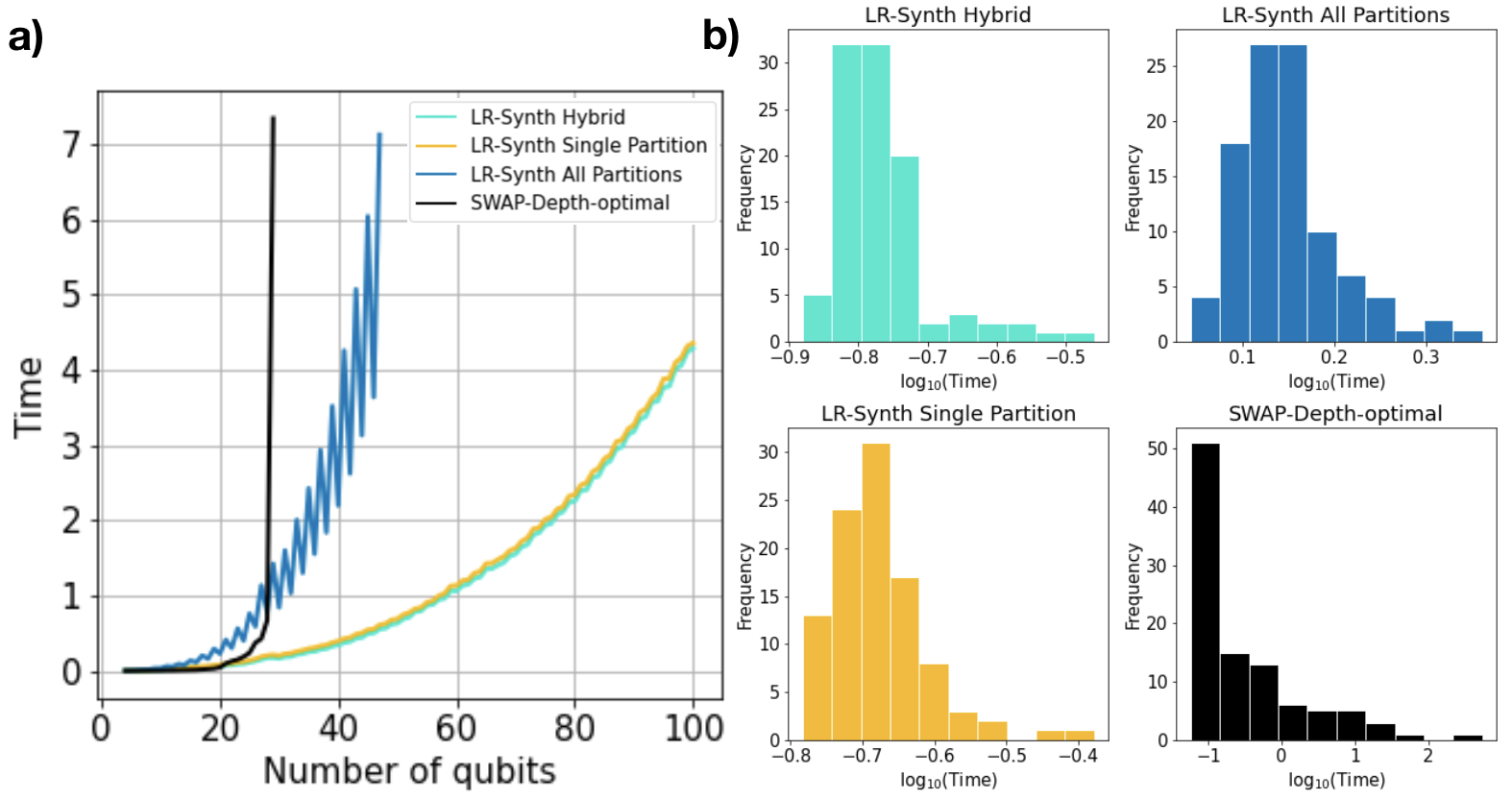}
 \caption{a) Scaling comparison of LR-Synth and Tweedledum SAT using a ring topology; the time on the Y axis is in arbitrary units since the algorithms were written in different languages and their runtimes cannot be directly compared. b) Tweedledum's SAT Solver spends substantial time on certain `hard' permutations starting at around 29 qubits for ring topologies, which significantly increases average runtimes, while LR-Synth spends approximately the same amount of time on each permutation for a given number of qubits.}
 \label{fig:runtime}
 \end{figure}

 We compare the runtime of LR-Synth with that of Tweedledum's SWAP-depth-optimal SAT solver, by timing the runs of LR-Synth sampling a single partition, LR-Synth sampling all partitions, and Tweedledum SAT, on a ring topology on a typical personal computer (See Figure \ref{fig:runtime}). We find that Tweedledum SAT works up to around 29 qubits for a ring topology, after which it spends hours trying to find a solution for certain permutations. Overall, sampling a single partition for LR-Synth is significantly faster than Tweedledum SAT or sampling all partitions for larger qubit numbers, and the performance of sampling a single partition is comparable to that of sampling all partitions (Figure \ref{fig:lr-synth}). 

 Note that while the scalability of LR-Synth and Tweedledum SAT can be compared, the actual runtimes are not directly comparable because LR-Synth is implemented in Python and is not optimized for classical runtime, as opposed to Tweedledum SAT's implementation in C++. 

\section{Application in Circuit Compilation}\label{sec:practical}

To demonstrate how our algorithms can be applied to compiling full circuits, we investigate the problem of compiling Quantum Volume circuits onto a path architecture.
We do this because prior to this point, the paper has focused on individual permutations --- but we would like to know whether easy or hard permutations occur in real circuits, and also how significant their optimization is to the overall circuit cost.

Quantum Volume (QV) is a holistic metric for benchmarking quantum computers~\cite{Cross}. It quantifies the largest random square circuit (i.e. equal logical width and depth) that a quantum computer can reliably execute. Reducing the size and/or depth of the circuit during compilation can translate to an improved quantum volume.

A QV circuit consists of several layers of random two-qubit unitaries applied to randomly chosen pairs of qubits. In the absence of all-to-all connectivity, this results in qubit communication requirements which are equivalent to permutations. This is typical in many other circuits of various applications as well. We seek to optimize these permutations using the methods presented in this paper. The outline of our compilation flow is as follows. 

\begin{enumerate}
    \item Use a standard mapping algorithm to embed the circuit onto the physical qubits and gates (here we use SabreLayout
    and StochasticSwap from Qiskit).
    \item Collapse each resulting network of SWAPs into a ``permutation" operation.  We use a greedy algorithm to grow a bag of SWAPs until we exhaust possible additions, and replacing that block with its equivalent permutation.
    \item Apply a permutation synthesis method on each permutation in the circuit. If the resulting synthesis is better than the original SWAP network under the considered cost function (size or depth), then the permutation is replaced by the new synthesis. If there is a tie, the secondary cost (depth or size) is considered.
    \item Translate the circuit to the target basis gates and optimize. Regular compilation techniques from the literature are applicable~\cite{jurcevic2021demonstration}.
\end{enumerate}

\begin{figure*}[htp]
\centering
\includegraphics[width=2\columnwidth]{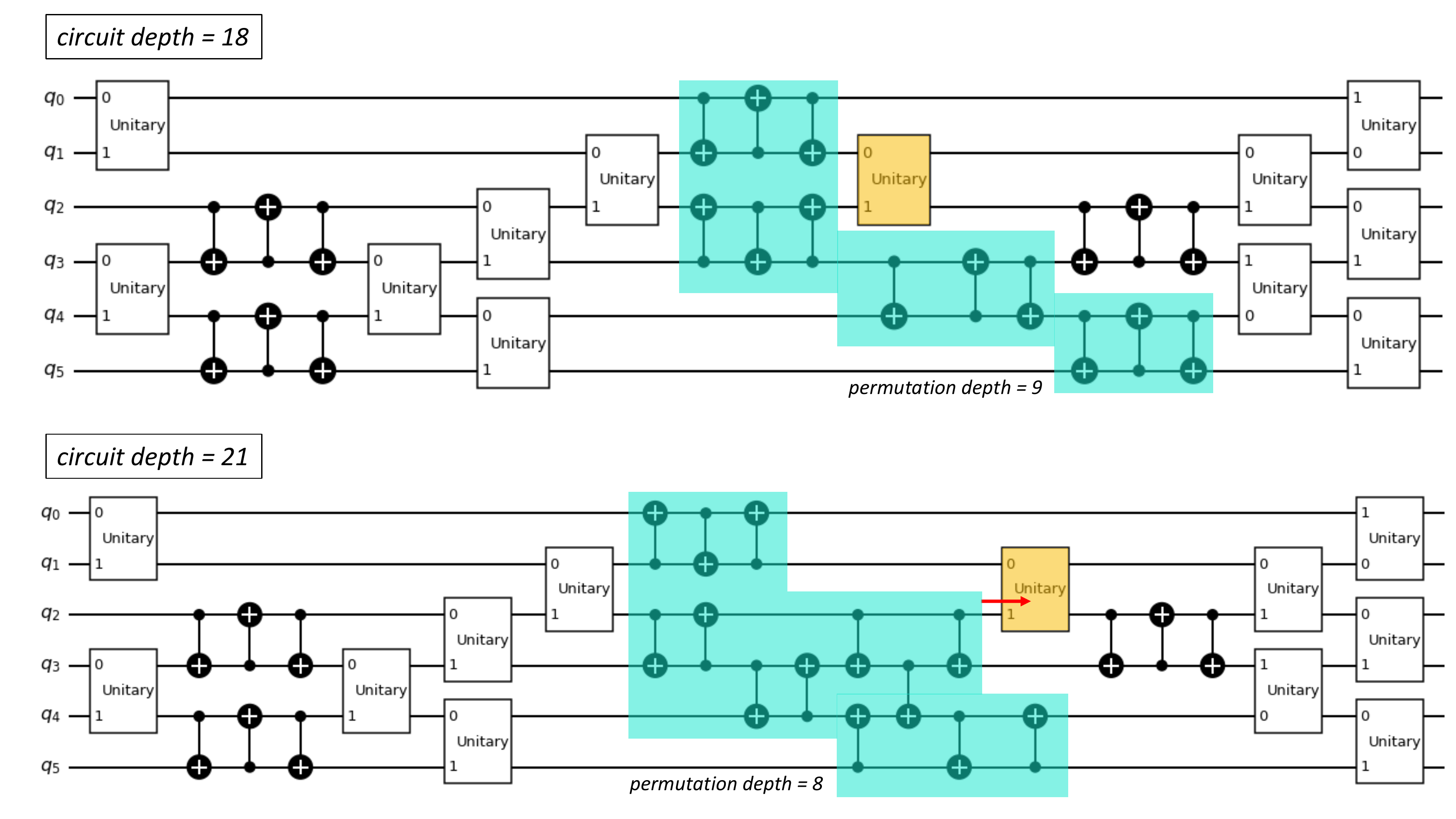}
\caption{Illustration of why local improvements to circuit depth do not necessarily improve global depth. The sub-circuit under the blue shade is a permutation, which can be re-synthesized for better depth. This alternate synthesis, however, shifts the highlighted unitary to the right. Since this lies on the circuit's critical path, it has ripple effects that increase the total circuit depth. Therefore, in addition to the depth of locally synthesized circuits, their {\bf shape} is also important for global depth.}
\label{fig:qv-circuit}
\end{figure*}

There are a few technical details in the above procedure which are worth elaborating. 
First, gathering all SWAPs in Step 2 may be harmful, since some of those SWAPs occur immediately after a two-qubit unitary. In effect, those SWAPs are free as they can be absorbed into the unitary (to yield a mirrored unitary) at no additional decomposition cost. But if those SWAPs are instead considered part of a permutation, then the resulting permutation synthesis may no longer result in such cost-saving absorption. Therefore, prior to Step 2, we consolidate any SWAPs that can be absorbed into an adjacent unitary. 

Second, depth optimization is not as straightforward as size optimization. In size optimization, any improvements during synthesis of individual permutations add up together. However, local reductions to depth during permutation synthesis do not necessarily translate to global reductions in the circuit depth, since that depends on how sub-circuits interact with each other. Therefore it is possible that some depth-oriented synthesis methods actually {\em increase} overall circuit depth. This is especially true when the synthesis method solely focuses on depth and neglects size, as those extra gates could adversely affect how the downstream gates slide on top of the synthesized circuit. We observe this to happen when using SWAP-depth-optimal circuits. A simple example illustrating this for a small circuit is shown in Figure~\ref{fig:qv-circuit}.

\begin{figure*}[htp]
\centering
\includegraphics[width=2\columnwidth]{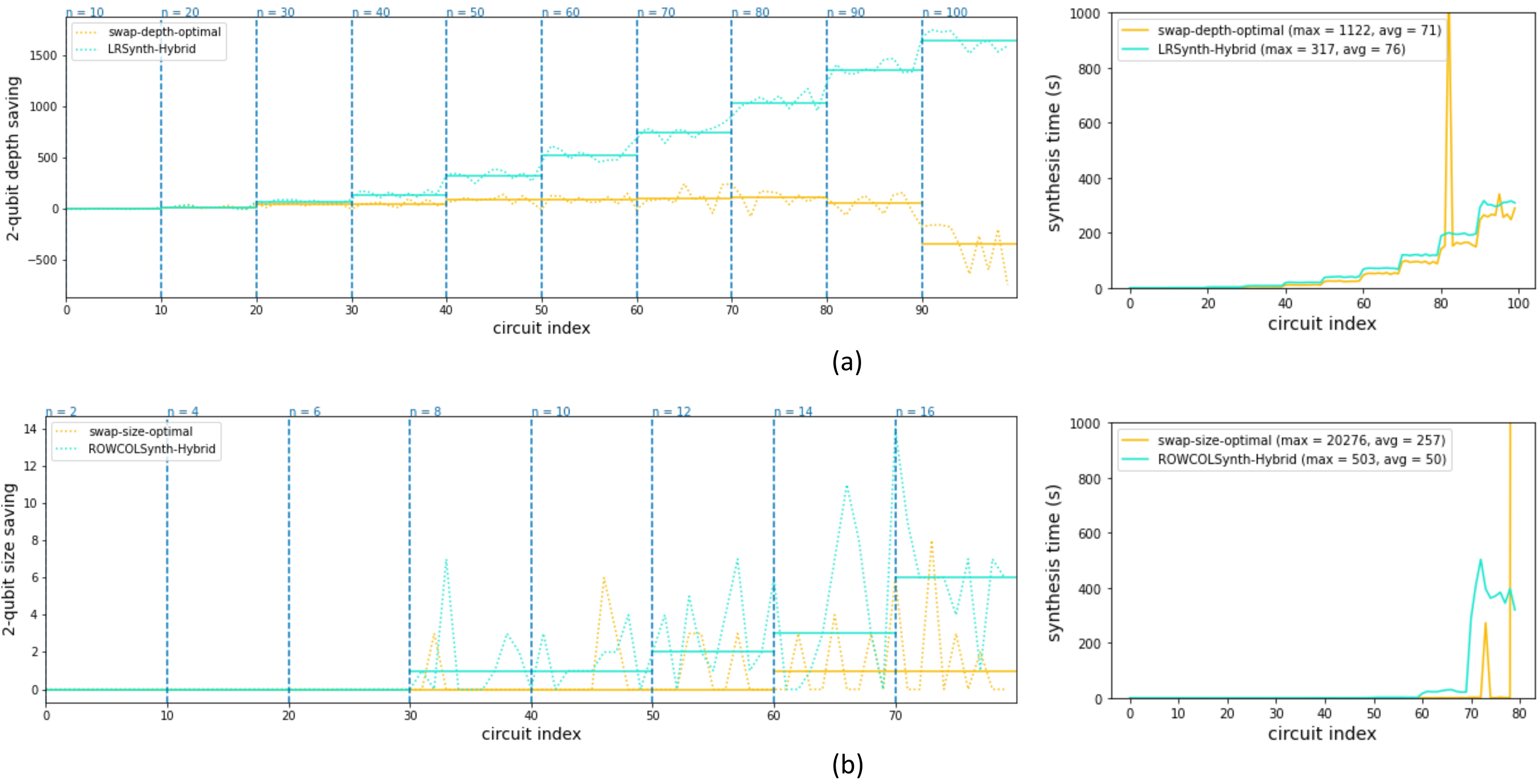}
\caption{Compilation of quantum volume circuits to a path architecture by re-synthesizing permutations, optimizing for (a) depth or (b) size. For each case the savings via exact and heuristic methods that optimize for that objective are shown. LR-Synth Hybrid and ROWCOL-Synth Hybrid outperform exact SWAP-depth and SWAP-size minimizers, respectively. In the former case, this is due to better synthesis size for the same depth. In the latter, the use of CNOTs improves the synthesis.
Illustrated on the right is the time to synthesize permutations in each circuit. Timing in heuristic methods increase predictably with permutation size. Exact solvers, while very fast for easy permutation, scale poorly when hard permutations are encountered.}
\label{fig:qv-savings}
\end{figure*}

The results of our depth-oriented and size-oriented compilations are depicted in Figure~\ref{fig:qv-savings}. We apply LR-Synth and SWAP-depth-optimal synthesis on QV circuits of up to 100 qubits on a path architecture. We further apply ROWCOL-Hybrid and SWAP-size-optimal synthesis on QV circuits of up to 16 qubits. LR-Synth is hybridized with SWAP-depth-optimal synthesis for $n \leq 20$ qubits. ROWCOL is hybridized with CNOT-size-optimal synthesis for $n \leq 4$ qubits. At every permutation encountered in the circuit, the synthesis is chosen only if it improves on the original synthesis under the objective of interest.

We observe that LR-Synth is able to outperform SWAP-depth-optimal synthesis substantially. This may be surprising since both methods are SWAP based, and the latter is optimal. Upon closer inspection, we can see that this is due to the particular shape of synthesis from both methods. In particular, the solver-based method has no regard for circuit size, and only minimizes depth. However there may be multiple solutions at the same depth. LR Synth is not only able to reach an optimal solution in a majority of cases, it chooses solutions that do not suffer from high size. This becomes crucial in the global depth, as bad size can have ripple effects on the global circuit depth, as illustrated in Figure~\ref{fig:qv-circuit}.

ROWCOL-Hybrid is also able to outperform the SWAP-size-optimal method. While local savings translate easily to global savings in the case of size, the advantage here comes from the fact that ROWCOL is CNOT-based, and can therefore find solutions inaccessible to even optimal SWAP-based methods.

Finally, we compare the timing between these methods. The absolute values of runtime are not the focus, since the implementations are in different languages and the heuristics have not been optimized for time. Nevertheless, we highlight a trend: heuristic methods such as LR-Synth and ROWCOL have a runtime that mainly depends on the number of qubits. That is, the specific permutations do not alter the timing significantly, which leads to a predictable step-like increase as larger circuits are considered. Exact methods, however, are mainly affected by the permutation under consideration. We see that many of the permutations that arise are ``easy'' and fast to synthesize. However, a few spikes are seen due to encountering hard permutations. These can take hours to synthesize, which leads to less predictable runtimes.

\section{Conclusion and Open Questions}
We proposed two algorithms for synthesis on limited-connectivity quantum computers. Using a modified ROWCOL algorithm hybridized with our optimal size solver for CNOTs, we showed that permutations synthesized using CNOTs can significantly outperform the optimal circuit sizes achievable by SWAP-based methods. It follows that an important open question is whether there is a depth-optimizing algorithm for general topologies using CNOTs and how much improvement it can yield. It would also be useful to improve the runtime of ROWCOL by having better heuristics for choosing the removal node at each step.

We proposed LR-Synth as a scalable depth-optimizing algorithm applicable to any topology. It achieves similar circuit depth as state-of-the-art heuristics tailored to paths, trees, and grids, and scales much better than optimal solvers. How much can LR-Synth be improved with better heuristics for partitioning the graph, choosing which path each qubit is routed through, or even increasing the number of partitions? 

We showed that global depth reduction is challenging since local reductions do not necessarily accumulate. Are there good heuristics for depth synthesis that is aware of the broader circuit context?

Finally, worst-case depth for permutation on a path remains an open question. We have shown a counter-example to the conjecture that reversal is the hardest permutation. What is the worst-case CNOT depth? We conjecture this to be $2n+O(1)$, rather than the known upper bound of $3n$~\cite{Kutin}.

\section{Acknowledgments}

This work was partially supported by the U.S. Department of Energy, Office of Science, National Quantum Information Science Research Centers, Co-design Center for Quantum Advantage (C2QA) under contract number DE-SC0012704

\FloatBarrier






\bibliographystyle{ACM-Reference-Format}
\bibliography{references}

\end{document}